\documentclass[12pt]{article}

\begin{document}

\author{A.I.Volokitin$^{1,2}$ and B.N.J.Persson$^1$ \\
\\
$^1$Institut f\"ur Festk\"orperforschung, Forschungszentrum \\
J\"ulich, D-52425, Germany\\
$^2$Samara State Technical University, 443100 Samara,\\
Russia}
\title{Adsorbate vibrational mode  enhancement of  radiative heat transfer}
\maketitle

\begin{abstract}
We show that  the radiative  
heat transfer between two  solid surfaces at short separation  may increase  by many order of
magnitude when the surfaces are covered by adsorbates.
 In this case
the heat transfer  is determined by resonant photon tunneling between adsorbate
vibrational modes. We propose an experiment to check the theory.
\end{abstract}

It is well known that for bodies separated by $d>>d_T= c\hbar /k_BT$, the
radiative heat transfer between them is described by the Stefan- Bolzman
law:
\begin{equation}
S=\frac{\pi ^2k_B^4}{60\hbar ^3c^2}\left( T_1^4-T_2^4\right)
, \label{stefan}
\end{equation}
where $T_1$ and $T_2$ are the temperatures of solid $\mathbf{1}$ and 
$\mathbf{2}$,
respectively. In this limiting case the heat transfer is connected with
traveling electromagnetic waves radiated by the bodies, and does not depend
on the separation $d$. For $d<d_T$ the heat transfer increases by many order
of magnitude due to the  evanescent
 electromagnetic waves  that decay exponentially into the vacuum; this is often refereed to as 
photon tunneling.  At present
 there is an increasing number of investigations of heat transfer due to
evanescent waves in connection with the scanning probe  microscopy
 under ultrahigh vacuum conditions \cite
{Van Hove,Levin1,Pendry,Majumdar,Volokitin2,Volokitin3,Mulet}. 
 It is now possible to measure extremely small amounts of heat transfer into small
 volumes \cite{Barnes}. STM can be used for local heating of the
surface, resulting in local desorption or decomposition of molecular
species, and this offer further possibilities for the STM to control local
chemistry at surface.

The efficiency of the radiative heat transfer  depends strongly on the dielectric 
properties of the media. In \cite{Pendry,Volokitin2,Volokitin3} it was shown  
the heat flux can be greatly enhanced if the conductivities of the material 
is chosen to maximize the heat flow due to photon tunneling. At room temperature 
the heat flow is maximal at conductivities corresponding to semi-metals. In fact, 
only a  thin film ($\sim10${\AA}) of a  high-resistivity material  is needed to maximize 
the heat flux \cite{Volokitin2}. Another enhancement mechanism of the radiative heat transfer can 
be connected with resonant photon tunneling between states localized on the 
different surfaces. Recently it was discovered that resonant photon tunneling between
surface plasmon modes give rise to extraordinary enhancement of the optical
transmission through sub-wavelength hole arrays \cite{Ebbesen}. The same
surface modes enhancement  can be expected  for the radiative heat transfer (and the 
van der Waals friction \cite{Volokitin}) if the
frequency of these modes is sufficiently low to be excited by thermal radiation.
At room temperature only the modes with frequencies below
$\sim 10^{13}s^{-1}$ can be excited.
 Recently,  enhancement of the heat 
transfer due to resonant photon tunneling between surface plasmon modes localized on 
the surfaces of the semiconductors was predicted  \cite{Mulet}.  
Other  surface modes which can be excited
by thermal radiation are
adsorbate vibrational modes.

 In this paper we study the  radiative heat
transfer between  small particles, e.g. adsorbed molecules, or
 dust particles, considered as  point dipoles, and
 situated on the surfaces of  different bodies. 
 Using an electromagnetic approach, in the dipolar approximation, we  derive a general 
expression the  the radiative heat power exchanged between the particles. We show that, 
if the particles have resonance frequencies which are matched, the heat transfer can
be enhanced by many orders of magnitude in comparison with the heat transfer between clean surfaces of the
good conductors.

Let us consider two particles with  dipole polarizabilities $\alpha_1(\omega)$ and 
$\alpha_2(\omega)$ and with the fluctuating dipole moments $p_1^f$ and $p_2^f$  normal 
to the surfaces. Accordingly 
to fluctuation-dissipation theorem \cite{Landau},  the spectral density function for the fluctuating dipole 
moment is given by
\begin{equation}
\langle p_i^fp_j^f\rangle _{\omega}=\frac{\hbar}{\pi}\left(\frac{1}{2}+n_i(\omega)\right)\mathrm{Im}
\alpha_i(\omega)\delta_{ij}
\label{one}
\end{equation}
where the Bose-Einstein factor 
\begin{equation}
n_i(\omega)=\frac{1}{e^{\hbar \omega/k_BT_i}-1}.
\end{equation}
Assume that the particles are situated opposite to each other on two different surfaces, at the  temperatures 
$T_1$ and $T_2$, respectively, and separated by the 
distance $d$. The fluctuating electric field of  a particle $\mathbf{1}$ does work on a particle $\mathbf{2}$.
The rate of working  is determined 
by
\begin{equation}
P_{12}=2\int_0^{\infty}d\omega\,\omega \mathrm{Im}\alpha_2(\omega)\langle E_{12}E_{12}\rangle  _{\omega}
\label{two}
\end{equation}
where  $E_{12}$ is the electric field created by a particle $\mathbf{1}$ at the position of a particle $\mathbf{2}$: 
\begin{equation}
E_{12}=\frac{8p_1^f/d^3}{1-\alpha_1 \alpha_2 (8/d^3)^2}
\label{three}
\end{equation}
From Eqs. (\ref{one}-\ref{two}) we get $P_{12}$, and  the rate of cooling of  a particle $\mathbf{2}$ 
can be obtained  
using the same formula by reciprocity. Thus the total heat exchange power  between the particles is given by
\begin{equation}
P=P_{12}- P_{21}=\frac{2\hbar}{\pi}\int_0^{\infty}d\omega\,\omega \frac{\mathrm{Im}\alpha_1\mathrm{Im}\alpha_2(8/d^3)^2}
{\left|1-(8/d^3)^2\alpha_1\alpha_2\right|^2} (n_1(\omega)-n_2(\omega))
\label{four}
\end{equation}

Let us firstly consider some general  consequences of Eq. (\ref{four}).
There are no constrain on the particle polarizability $\alpha(\omega)=\alpha^{\prime}+i\alpha^{\prime \prime}$ 
 other than that $\alpha^{\prime \prime}$ is positive, and $\alpha^{\prime}$ and $\alpha^{\prime \prime}$ are connected by the 
Kramers-Kronig 
relation. Therefore, assuming identical surfaces, we are free to maximize the photon-tunneling 
transmission coefficient  
\begin{equation}
t=\frac{(8\alpha^{\prime \prime}/d^3)^2}{\left|1-(8\alpha /d^3)^2\right|^2}
\label{five}
\end{equation}
This function is a maximum when 
\begin{equation}
\alpha^{\prime 2} + \alpha^{\prime \prime 2} =(d^3/8)^2
\end{equation}
so that $t=1/4$. Substituting this result in (\ref{four}) gives the upper bound for 
the heat transfer power  between two particles   
\begin{equation}
P_{max}=\frac{\pi k_B^2}{3\hbar}(T_1^2-T_2^2)
\label{six}
\end{equation}
 For adsorbed molecules at the  concentration $n_a=10^{19}$m$^{-2}$, when one surface is at zero temperatures 
and the other is at the room temperature the maximal heat flux due to the adsorbates $S_{max}=n_aP_{max}=10^{12}$Wm$^{-2}$, 
which is nearly 10 order of magnitude larger than the  heat flux due to the black body radiation,
 $S_{BB}=\sigma T =4\cdot 10^2$Wm$^{-2}$.

We rewrite the denominator of the integrand  in Eq. (\ref{four})
 in the form
\[
\left|1-(8\alpha/d^3)^2\right|^2=[(1-8\alpha^{\prime}/d^3)^2+(8\alpha^{\prime \prime}/d^3)^2]
\]
\begin{equation}
\times[(1+8\alpha^{\prime}/d^3)^2+(8\alpha^{\prime \prime}/d^3)^2]
\label{seven}
\end{equation}
The conditions for resonant photon tunneling are determined by equation
\begin{equation}
\alpha^{\prime}(\omega_{\pm})=\pm d^3/8
\label{eight}
\end{equation}   
Close to resonance we can write
\[ 
(1\pm 8\alpha^{\prime}/d^3)^2 +  (8\alpha^{\prime \prime}/d^3)^2
\]
\begin{equation}
\approx 
(8\beta_{\pm}/d^3)^2[(\omega-\omega_{\pm})^2+((\alpha^{\prime \prime}/\beta_{\pm})^2]
\label{nine}
\end{equation}
where
\[
\beta_{\pm}=\left.\frac{d\alpha_r^{\prime}(\omega)}{d\omega}
\right|_{\omega=\omega_{\pm}},
\]
Assuming $|\alpha^{\prime \prime}/\beta_{\pm}|<<\omega_{\pm}$  we get the following contribution to  the heat 
transfer:
\begin{equation}
P=\frac{\hbar}{2}[(\alpha^{\prime \prime}(\omega_{+})/|\beta_{+}|)\omega_{+}(n_1(\omega_+)-n_2(\omega_+)) 
+ (+\rightarrow -)]
\label{ten}
\end{equation}
 Close to a  pole we can  
 use the approximation
\begin{equation}
\alpha\approx\frac{a}{\omega-\omega_0-i\eta},
\label{eleven}
\end{equation}  
where $a$ is a constant. Then from the resonant condition (\ref{two}) we get
\[ 
\omega_{\pm}=\omega_0\pm 8a/d^3.
\]
For the two- poles approximation to be valid the difference $\Delta\omega =
|\omega_+-\omega_-|$ must be greater than the width $\eta$ of the resonance, 
 so that 
$8a/d^3>\eta$.

For $\eta<<8a/d^3$,  from Eq. (\ref{four}) we get
\begin{equation}
P=\frac{\hbar \eta}{2} [\omega_+ (n_1(\omega_+)-n_2(\omega_+) + (+\rightarrow -)].
\label{twelve},
\end{equation}
Using Eq. (\ref{twelve}) we can estimate the heat flux between identical surfaces covered by  adsorbates with concentration $n_a$: 
 $S\approx n_a P$.
Interesting, the  explicit $d$ dependence has dropped out of Eq. (\ref{twelve}). 
However, $P$ may still be $d$- dependent, through the $d$- dependence 
of $\omega_{\pm}$. For  $\hbar\omega_{\pm}\le k_BT$ 
 the heat transfer  will be only  weakly distance independent. 

For $8a/d^3<\eta$ we can neglect  multiple scattering of the photon between the particles, 
so that   the denominator in the integrand in Eq. (\ref{four}) equal unity. For $d>>l$, where 
$l$ is the interparticle spacing, the heat flux between two surfaces covered by adsorbates 
with concentration $n_{a1}$ and $n_{a2}$ can be obtained after integration of the heat flux between 
two separated particles. We get
\begin{equation}
S=\frac{24\hbar n_{a1}n_{a2}}{d^4}\int_0^{\infty}d\omega\,\omega \mathrm{Im}\alpha_1 \mathrm{Im}\alpha_2
[n_1(\omega)-n_2(\omega)]
\label{thirteen}
\end{equation}
Assuming that $\alpha$ can be approximated by Eq. (\ref{eleven}),   for $\omega_0<<\eta$  Eq. (\ref{thirteen}) 
gives  the heat flux between two identical surface:
\begin{equation}
S=\frac{12\pi \hbar \omega_0 a^2n_a^2}{d^4\eta}[n_1(\omega_0)-n_2(\omega_0)]
\label{fourteen}
\end{equation}

 In the case of  ionic adsorption the dipole polarazibility is given by
\begin{equation}
\alpha=\frac{e^{*2}/M}{\omega^2-\omega^2_0-2i\omega\eta}
\label{fifteen}
\end{equation}
where $e^*$ is the ionic charge, M is the ionic mass, and where $\omega_0$ and $\eta$ are the vibrational 
frequency and damping constant, respectively. 
For the K/Cu(001) system $\omega_0=1.9\cdot10^{13}$, and at low coverage $e^{*}=0.88$ \cite{Senet}, 
which gives  $a=e^{*2}/2M\omega_0=7\cdot10^{-17}$
m$^3$s$^{-1}$. 
For $\eta = 10^{12}$s$^{-1}$ and $d<10${\AA}, when one surface has
$T=300$K and the other $T=0$K,  we get distance independent  
$P\approx10^{-9}$W 
. In this case,  for  $n_a=10^{18}$m$^{-3}$ 
 the heat flux $S\approx10^{9}$Wm$^{-2}$. At the same conditions the heat flux between two clean surfaces 
$S_{clean}\approx10^{6}$Wm$^{-2}$. Thus the photon tunneling between the adsorbate vibrational states can 
strongly enhance the radiative heat transfer between the surfaces.

Let us describe the physical origin of the different regimes in  resonant photon tunneling between 
adsorbate vibrational modes.
At sufficiently small separation, when  $8a/d^3>\eta$, the photons goes back and forth several time
in the vacuum gap, building up coherent constructive interference in the forward direction
much as would occur in resonant electron  tunneling. In this case the vibrational modes  on the isolated
surfaces combine to form  collective vibrational modes  (diatomic "molecule"), where 
  the adsorbates vibrate in phase or
 out of phase.
 This will result in a  very weak distance
dependence of the heat flux, because the transmission probability for photon   depends 
very weakly on $d$ in this case
(see above). For large $d$, when  $8a/d^3<\eta$, the  sequential tunneling is more likely to occur, where the
photon excited in a adsorbate vibrational mode, tunnels to the adsorbate vibration  at the other surface, and then
couples to the other excitations in the media and exit.

The above discussion is for a  special case of matching adsorbate vibrational frequencies ($\omega_1=\omega_2 =
\omega_0$), but the picture still applies in the nonsymmetric case ($\omega_1\neq\omega_2$). Here, adsorbate vibrational 
modes on the two surfaces have different frequencies and, instead of a vibration mode for diatomic homopolar 
 molecule, we have a 
diatomic heteropolar molecule.

Finally, let us suggest an experiment to probe the photon-tunneling heat transfer theory. Consider a solid surface 
(substrate) at low 
temperature with a low concentration of weakly adsorbed atoms (e.g. noble gas atoms) or molecules. The position of the atoms can 
mapped out relative to the substrate using a STM. Next, the surface of another solid at higher temperature (e.g. room temperature) 
is brought in the vicinity (separation $d$) of the substrate. The heat transfer (via photon tunneling) to the substrate will 
result in a temperature increase on the substrate surface. This will  result in the diffusion of the weakly bound adsorbates. 
The (average)  diffusion distance will be a function of the heat transfer. If the two bodies are separated after a given time period, 
and if the new position of the adsorbates is determined   using the STM  \cite{Briner}, it is possible to infer the heat transfer from 
the hot to the cold surface as a function of the separation $d$.  This experiment is conveniently performed using as the second (warm) body an AFM tip with a wide flat tip, as recently produced and used for contact mechanics studies \cite{Buzio}.

In summary, we have presented a detailed theoretical study of of the heat transfer between surfaces covered by adsorbates. 
We have shown that resonant photon tunneling between adsorbate vibration modes 
can give rise extraordinary  enhanced heat transfer,  in the comparison with the heat transfer between clean 
good conductor surfaces. This result can be used in the scanning
probe microscopy for local heating and modification of the surface. Finally, we have  suggested 
 an experiment 
by which the radiative heat transfer due to photon tunneling can be measured.

\vskip 0.5cm \textbf{Acknowledgment }

A.I.V acknowledges financial support from DFG 
, B.N.J.P.  acknowledges 
support from the European Union Smart Quasicrystals project.

\thinspace \thinspace \thinspace

\end{document}